# Computer aided planning for orthognatic surgery

Matthieu Chabanas [a], Christophe Marécaux [a,b], Yohan Payan [a], Franck Boutault [b]

[a] TIMC-IMAG Laboratory, Université Joseph Fourier, Grenoble, France
Institut Albert Bonniot, 38706 La Tronche cedex, France
[b] Department of maxillofacial and facial plastic surgery – CCRAO Laboratory,
Hôpital Purpan, 31059 Toulouse cedex, France

## Abstract

A computer aided maxillofacial sequence is presented, applied to orthognatic surgery. It consists of 5 main stages: data acquisition and integration, surgical planning, surgical simulation, and per operative assistance. The planning and simulation steps are then addressed in a way that is clinically relevant. First concepts toward a 3D cephalometry are presented for a morphological analysis, surgical planning, and bone and soft tissue simulation. The aesthetic surgical outcomes of bone repositioning are studied with a biomechanical Finite Element soft tissue model.

## 1. Introduction

Orthognatic surgery, as a part of cranio-maxillofacial surgery, attempts to establish normal functional and aesthetic anatomy for patients suffering from dentofacial disharmony, by repositioning maxillary and mandibular osteotomies.

The current therapeutic protocol is a difficult and laborious process that might be responsible for inaccuracy in comparison with therapeutic planning. It is now well accepted that medical imaging and computer assisted surgical technologies may improve current orthognatic protocol as an aid in diagnostic, surgical planning and surgical intervention. The sequences required for a complete computer assisted protocol in cranio-maxillofacial surgery are well described in the literature since the eighties. Although, in orthognatic clinical practice this protocol still fails on diagnostic and planning in a three-dimensional environment. There is neither three-dimensional skeletal analysis for morphological diagnostic and osteotomy simulation (called 3D cephalometry), nor reliable prediction of post operative facial appearance. This last point is important for surgical simulation as the final soft tissue facial appearance might modify the operative planning. Therefore, the patient also expects a reliable prediction of his post operative aesthetic appearance.

After reminding the sequences of a computer aided cranio-maxillofacial protocol, our own process in computer assisted orthognatic surgery is presented.

## 2. Method

### 2.1. Computer assisted cranio-maxillofacial surgery
The different steps and specifications of a computer aided protocol in cranio-maxillofacial surgery are well defined the in literature [1,2,3]. They can be summarized as in figure 1.

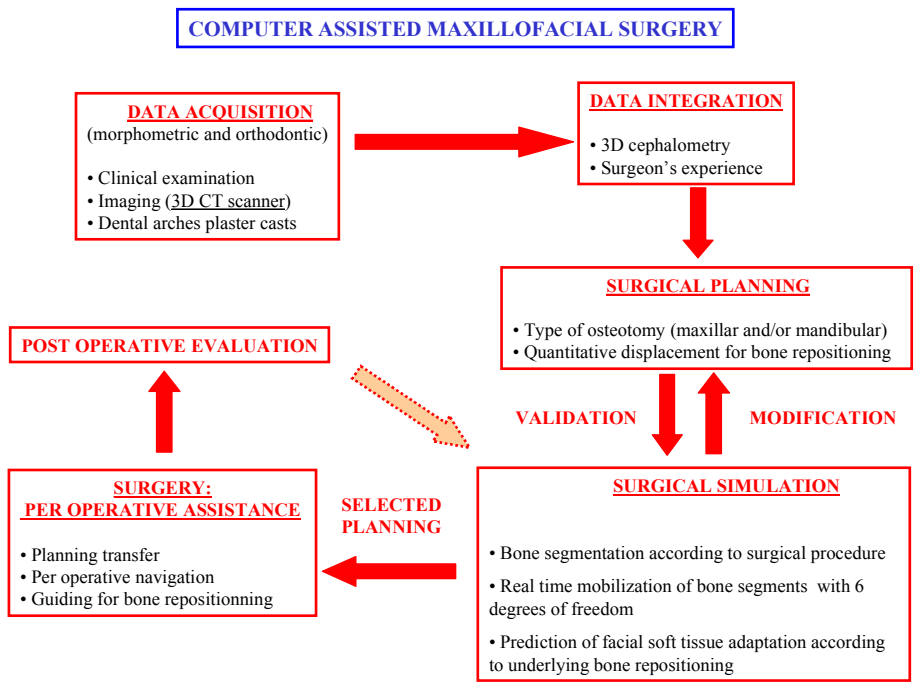

Figure 1. Specifications of a cranio-maxillofacial protocol. The dashed arrow concerns research context only: the post operative evaluation is used for the surgical simulation validation

Different parts of this protocol have been addressed in the literature. Most of them deal with interactive simulation of a surgical procedure on 3D computer generated models and present interesting tools to cut and manipulate bone segments [4,5,6]. However, none of them integrate morphometric and orthodontic data (cephalometric analysis) to establish a surgical planning, and are therefore relevant for clinical practice.

Three-dimensional cephalometric analysis, despite being essential for clinical use of computer aided techniques in maxillofacial surgery, has been studied very little so far. Most of these previous work [1,2], including our group [7], have presented extensions of 2D cephalometry. Cephalometric and orthodontic diagnostic and planning were made in traditional way (on standard 2D teleradiography and plaster dental casts) or with three dimensional constructions from 2D data. One of the most original work has been presented by Treil [8], who introduced a new cephalometry based on CT scanner imaging, anatomic landmarks and mathematic tools (a maxillofacial framework and dental axis of inertia) for skeletal and dental analysis. However, in our point of view, this cephalometric analysis is not relevant for surgical planning and computer guided surgery.

Physical models were also developed to evaluate the aesthetic outcomes resulting from underlying bone repositioning [4,9,10,11]. They commonly use a continuous model based on the Finite Element method, with noticeable differences. According to us, despite their evident scientific interest, most of these works cannot be used in clinical practice since the bone simulation is not clinically relevant and the model generation is highly time consuming.

Our process in computer aided orthognatic simulation emphasises on future applications in a complete clinical protocol, as described in figure 1. In this way, we are developing a facial skeletal model including 3D cephalometry and a Finite Element model of the facial soft tissue for surgical planning and simulation issues.

## 2.2. Three dimensional cephalometry

A complete computer aided cranio-maxillofacial surgery sequence requires a craniofacial model that enables the morphological diagnostic, supports the surgical bone simulation, integrates the prediction of post operative facial soft tissue deformation, and can be used as interface in computer guided surgical stage.

To be accepted by medical community, this model must be coherent from an anatomical, physiological and organ genetic point of view. A 3D cephalometric tool as an aid in diagnostic is admitted as necessary [1,2,7]. 3D CT scanner imaging is already currently used to apprehend the difficult three dimensional part of this pathology. However, there is no relevant direct three dimensional analysis method of these images (3D cephalometry).

A reliable cephalometry requires to define a referential for facial skeleton orientation, used for intra and inter patient measurements reproducibility and for quantification of bone displacements, and a facial morphologic analysis method for treatment planning decision in comparison to a norm determined as "equilibrated" face.

This model should be able to be cut for simulation as in a surgical procedure. The Finite Element facial soft tissue model subsequently described in section 2.3 should also be integrated.

### 2.2.1. Referential definition

An invariant, reproducible, orthogonal referential composed of 3 planes is proposed (figure 2 left). An horizontal plane is defined, close to the cranio basal planes of previous 2D cephalometries and to the horizontal vestibular plane defined as the craniofacial physiological horizontal plane. Its construction use anatomic reliable landmarks: right and left *capitus mallei* and the middle point between both *foramen supraorbitale*. The medial sagittal and frontal planes are orthogonal to the horizontal one and contain the middle point of both *capitus mallei*. As defined, this referential is independent from the analysed facial skeleton, and is not modified by the surgical procedure.

The x, y and z coordinates of each voxel are transferred from the original CT scanner referential to this new referential. These normalized coordinates allow location or measurement comparison between two patients or in the same one across time.

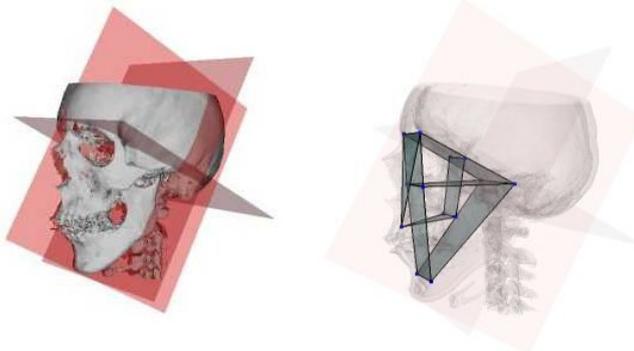

Figure 2. Craniofacial referential (left) and a maxillofacial framework example with anterior and posterior facial surfaces, prefacial surface, upper medium and lower facial surfaces and the palatine surface (right)

## 2.2.2. Maxillofacial framework for skull analysis

The cephalometric definition requires both a maxillofacial framework for morphologic analysis and a norm defined as an ideal for a pleasant equilibrated face. The operative planning is defined by difference between the current patient state and the norm. This one should be defined as average numeric values in location of special landmarks or as an equilibrated construction.

A maxillofacial framework is presented, composed of 15 anatomic reliable landmarks (*capitus mallei, foramen supraorbitale, foramen infraorbitale, foramen mandibulae, foramen mentale, foramen palatinum majus* and *foramen rotundum* on each side and the *foramen incisivum*) and 8 surfaces (upper and lower anterior facial s., prefacial s., posterior facial s., upper facial (cranio basal) s., medium facial s., lower facial s. and palatine s.). An example of construction is shown in figure 2 (right).

Mathematical tools allow metric, angular and surfacic measurements. Unlike traditional 2D cephalometry, these are direct values and not measurements between projected and constructed points on a sagittal radiography.

## 2.3. Finite element model of the face soft tissue

Different face models have been developed for simulating the aesthetic outcomes of maxillofacial surgery. Although the first ones were based on discrete mass-spring structures [9], most of them use the Finite Element method [4,10,11] to resolve the mechanical equations describing soft tissue behaviour. These methods are based on a 3D volumetric mesh, generated from patient CT images using automatic meshing methods. Such algorithms are not straightforward to use in this case, as the boundary of the face tissue, i.e. the skin and the skull surfaces, must be semi-automatically segmented, which is highly time-consuming and cannot be used in current clinical activity. Moreover, these meshes are composed of tetrahedral elements, usually less efficient than hexahedral ones in terms of accuracy and convergence.

## 2.3.1. Patient specific mesh generation

Our method [12] is, first, to build one "generic" model of the face composed of 2 layers of hexahedral elements representing dermis and hypodermis. As elements are structured within the mesh, main mimic muscles are modelled.

Then, this generic mesh is adapted to each patient morphology, built out of CT scanner data, using a mesh-matching algorithm. This method, based on the octree spline elastic registration algorithm, computes a non-rigid transformation between 3D surfaces, external nodes of the generic model and patient skin surface on one part, and internal nodes of the generic model and patient skull surface on the other part. Elements of the patient mesh are automatically regularised to enable Finite Element computation.

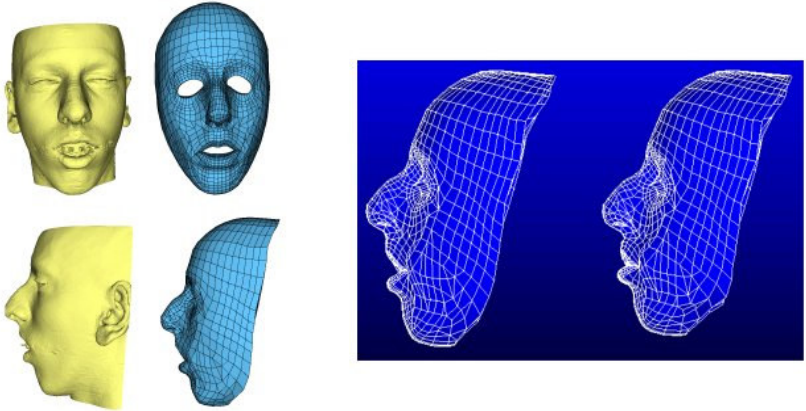

Figure 3. A patient specific finite elements model built from a CT exam (left) and simulation of soft tissue deformation resulting from a bimaxillary osteotomy on this patient

### 2.3.2. Mechanical properties

In a first approximation, a linear elasticity behaviour is assumed for facial tissues, with a small displacement hypothesis. The facial anisotropy is taken into account by setting linear transverse elasticity in the muscles fibres directions.

## 3. Results

The 3D cephalometric method allows direct three dimensional morphometrical measurements on CT scanner imaging for patient study. In the same normalized referential, bone displacements from the pathological state to the normalized predicted one are simulated as in the surgical procedure. The displacements of the anatomical landmarks used for the cephalometry are applied to the bone segments where they are located.

The mesh generation method was successfully used to build several patient models with different morphology. The total reconstruction of a patient model is carried out in about 15 minutes. The accuracy given by the matching algorithm is under 1mm. To simulate the aesthetic outcomes of bone repositioning, internal nodes rigidly fixed to the maxilla and mandible are displaced according to the surgical planning. A patient model is presented in figure 3, with a simulation of soft tissue adaptation according to mandible and maxilla repositioning.

## 4. Discussion

If these models get closer to a clinical application of computer assisted techniques in orthognatic surgery, we are aware of remaining problems.

The presented 3D cephalometric model integrates neither basilar mandibular ridge nor gonial angle studies nor dental analysis, which are clinically important. Neither is solved the morphometric norm problem. As no tool for morphometric measurements on 3D imaging exists yet (except Treil's one [8]), no normative data set is available.

The soft tissue model generation is an easy to use, straightforward, almost automatic and fast method. Hence, it is suitable to be used by a surgeon in the current planning elaboration. However, clinical quantitative study must be carried out to validate or modify functional behaviour of the model. It requires the comparison of simulated predictions with real surgical outcomes of the procedure, which has never been done in the literature. These works are currently under development.